\documentclass[aps,preprint,superscriptaddress,nofootinbib]{revtex4}
\usepackage[utf8]{inputenc}
\usepackage{amsmath,amssymb,bbm}
\usepackage{dsfont}
\usepackage{graphicx}
\usepackage{xcolor,slashed}
\pdfoutput=1
 \usepackage{appendix}

 \usepackage{hyperref}       % hyperlinks
 \usepackage{url}            % simple URL typesetting
 \usepackage{booktabs}       % professional-quality tables
 \usepackage{amsfonts}       % blackboard math symbols
 \usepackage{nicefrac}       % compact symbols for 1/2, etc.
 \usepackage{microtype}      % microtypography
 \usepackage{lipsum}

\begin{document}

%\title{Dark matter in the general 2HDM solution to the baryon asymmetry of the universe}
%\title{Dark matter in the general 2HDM top-driven Electroweak Baryogenesis scenario}
\title{Dark Matter in the General 2HDM \\ with Extra Top Yukawa Couplings}
\author{Leon M.G. de la Vega}
\email[E-mail: ]{leon.garcia@unison.mx}
\affiliation{Departamento de Física, Facultad Interdisciplinaria de Ciencias Exactas y Naturales, Universidad de Sonora, Hermosillo 83000, Mexico}
\affiliation{Physics Division, National Center for Theoretical Sciences, Taipei 106319, Taiwan}
\author{Mohamed Krab}
\email[E-mail: ]
{mkrab@hep1.phys.ntu.edu.tw}
\affiliation{Department of Physics, National Taiwan University, Taipei 106319, Taiwan}

\begin{abstract}
We extend the general two Higgs doublet model (G2HDM), which introduces extra Yukawa couplings, by an additional real scalar singlet (G2HDM+S), providing a viable scalar dark matter~(DM) candidate. This setup provide a viable UV completion of top-window dark matter scenarios, in which DM communicates with the standard model predominantly through the top quark. We analyze the constraints on the visible scalar sector from Higgs signal strength measurements and direct searches for additional Higgs bosons at the LHC, and those on the dark 
sector from cosmological observables, including the observed DM relic density and spin-independent direct-detection as well as indirect detection. Within the surviving parameter space, we identify a rich and diverse LHC phenomenology governed by the interplay between the singlet portal, the extended Yukawa sector, and the heavy scalar mass hierarchy. To facilitate experimental investigation, we propose six benchmark scenarios spanning the singlet-portal, bosonic-cascade, fermiophilic, and flavor-violating regimes. In the latter, the nondiagonal top-charm coupling $\rho_{tc}$ induces the production channel $cg \to tH$, allowing the top quark to serve as a trigger for an otherwise invisible $H\to SS$ signal. This flavor-violating DM production is a distinct feature of the G2HDM+S.
\end{abstract}

\maketitle

% keywords can be removed

%###################################################################
\section{Introduction}
The fundamental nature of dark matter~(DM) is one of the most pressing unanswered problems of modern physics. There is ample evidence that the Universe contains cold, non-luminous matter amounting to about 27$\%$ of its total energy density \cite{Rubin:1980zd,Clowe:2006eq,Planck:2018vyg}. The identification of the particle nature of dark matter is an active direction in experimental particle physics, in colliders, cosmic ray observatories and underground facilities. One of the best studied candidates for particle dark matter is the so-called weakly interacting massive particle~(WIMP), which is defined by its mass range from MeV to TeV, with weak-order couplings that lead to its thermalization with the standard model~(SM) bath in the early Universe.
Another gap in our understanding of fundamental physics is the origin of the observed baryon asymmetry of the Universe, which, according to the Sakharov conditions, calls for out-of-equilibrium dynamics, baryon number violating physics, and both C-symmetry and CP-symmetry violations~\cite{Sakharov:1967dj}. It is well known that although the SM contains these ingredients, it appears they are not enough to explain the observed value of the baryon-to-photon ratio $\eta_\gamma$. 
It has been shown that the elegantly simple extension of the SM consisting of a second Higgs doublet with no $Z_2$~symmetry, known as type-III two Higgs doublet model (2HDM) or general 2HDM~(G2HDM), can lead to phenomenologically viable electroweak baryogenesis~(EWBG)~\cite{Fuyuto:2017ewj}. This can be, for example, achieved through either a flavor-diagonal top coupling or a flavor-nondiagonal top-charm coupling. This model for EWBG can accommodate sub-TeV exotic scalars that can be searched for at the large hadron collider~(LHC) \cite{Kohda:2017fkn,Ghosh:2019exx,Hou:2024bzh}, as well as observably large electron electric dipole moment~(eEDM) and neutron EDM~(nEDM)~\cite{Fuyuto:2019svr,Hou:2021zqq,Hou:2023kho}. 

Motivated by these studies, we extend the G2HDM by a scalar DM candidate $S$, which we refer to as the G2HDM+S model.
Dark matter scenarios embedded in models with two active (non-inert) Higgs doublet models have been previously studied~\cite{Bell:2017rgi,Cabrera:2020lmg,Arcadi:2020gge,Arcadi:2024ukq}. In these scenarios, the dark and visible sectors communicate through the enlarged scalar sector, providing a richer dark matter phenomenology which can decouple the dark matter direct detection cross section from its annihilation cross section, thus solving the main drawback of models where only the SM Higgs boson acts as the portal between sectors~\cite{GAMBIT:2017gge,EscuderoAbenza:2025cfj}. 
In the case where natural flavor conservation (NFC) is not enforced, such as in the G2HDM realization, experimental constraints allow only hierarchical extra Yukawa couplings, where fermions of larger masses can have larger couplings to the second Higgs doublet~\cite{Crivellin:2013wna}. This suggests that in this framework, introducing a dark matter candidate can lead to phenomenology where the dark sector couples predominantly to top quarks and the additional scalars. 
Scenarios where dark matter communicates strongly with third-generation fermions have been found to be phenomenologically viable \cite{Cheung:2010zf,Demetriou:2025ewa}. In scenarios
which assume that the sector mediating interactions between the top and DM has been integrated out at the time of DM freezeout ($2 m_{\rm DM} \ll m_{\rm med}$), the possibilities of resonant annihilation, or new physics (beyond the DM candidate) are precluded.
In this work we study the G2HDM+S, an extension of the G2HDM by a real scalar singlet which plays the role of a scalar dark matter candidate. This setup provides a viable UV completion of the top-window scenarios of DM. The additional heavy scalars of the G2HDM inevitably enter the DM phenomenology in a non-trivial way: they modify the singlet relic density through new annihilation channels, shift the direct-detection predictions via mixing effects, and open new collider signatures that interleave DM missing-energy signals with Higgs-to-Higgs 
bosonic cascades and flavor-violating final states. 

This paper is organized as follows. We first introduce the G2HDM+S model in section~\ref{sec:model}. 
In section~\ref{sec:vispheno}, we discuss the constraints on the visible scalar sector of the model from Higgs signal strength measurements and direct searches for additional Higgs bosons at the LHC. We select benchmark points of scalar masses and mixings compatible with these constraints to study DM phenomenology.
We then discuss the constraints on the interactions to the dark sector from dark matter observables in section~\ref{sec:DMpheno}.
Next, we present in section~\ref{sec:BPs} some benchmark scenarios compatible with both scalar and dark sector observables for further experimental investigations.
We give our conclusions and outlook in section~\ref{sec:concl}.

%###################################################################
\section{G2HDM+S} \label{sec:model}
We consider the scalar extension of the SM known as the type-III 2HDM or G2HDM, with an additional real singlet, which will act as the dark matter candidate. For the stability of dark matter, one introduces an {\it ad~hoc} dark $Z_2^{D}$ symmetry acting non-trivially only on the scalar singlet~(see Table~\ref{tab:Z2D}).
\begin{table}[t]
    \centering
    \begin{tabular}{c|c|c|c}
       & $SU(2)_L$ & $U(1)_Y$ & $Z_2^{D}$ \\ \hline
        $H_1$ & 2 & 1/2 & + \\ 
        $H_2$ & 2 & 1/2 & + \\ 
        $S$ & 1 & 0 & - \\ 
    \end{tabular}
    \caption{Higgs fields under $Z_2^{D}$ symmetry.}
    \label{tab:Z2D}
\end{table}
With this matter and symmetry content, we adopt the Higgs basis, where only $H_1$ acquires a non-zero vacuum expectation value (vev) $v$. In this basis we have \cite{Davidson:2005cw}
\begin{equation}
    H_1= \begin{pmatrix}
          G^+\\
          \frac{1}{\sqrt{2}}(v + h_1 + i G^0 )
    \end{pmatrix}\; ,\;
    H_2= \begin{pmatrix}
          H^+\\
          \frac{1}{\sqrt{2}}( h_2 + i A )
    \end{pmatrix},
\end{equation}
with $G^+$ and $G^0$ being the $W$ and $Z$ boson Goldstone modes respectively.
 The scalar potential of this model, assuming CP invariance, is then
 \begin{equation}
     \begin{split}
         V&= \mu_{11}^2 H_1^\dagger H_1 + \mu_{22}^2 H_2^\dagger H_2 + \mu_S^2 S^2 + (\mu_{12}^2 H_2^\dagger H_1 +\text{h.c.})  \\
         &\quad + \frac{\eta_1}{2} (H_1^\dagger H_1)^2 + \frac{\eta_2}{2} (H_2^\dagger H_2)^2+  \eta_3 (H_1^\dagger H_1)(H_2^\dagger H_2)  + \eta_4 (H_1^\dagger H_2)(H_2^\dagger H_1)\\
         &\quad +\left(\frac{\eta_5}{2} (H_1^\dagger H_2)(H_1^\dagger H_2)+ \eta_6 (H_1^\dagger H_1)(H_1^\dagger H_2) + \eta_7 (H_2^\dagger H_2)(H_1^\dagger H_2) +\text{H.c.} \right)\\
        &\quad + \lambda_{11}S^2(H_1^\dagger H_1)+ \lambda_{22}S^2(H_2^\dagger H_2)+ \lambda_{12}S^2(H_1^\dagger H_2 + H_2^\dagger H_1),
     \end{split}
 \end{equation}
where $\eta_{1-7}$, $\lambda_{11}$, $\lambda_{22}$, and $\lambda_{12}$ are the quartic Higgs couplings and are real.
In the Higgs basis, the Yukawa couplings of $H_1$ give rise to fermion masses, so we can directly write the Yukawa interaction terms in the fermion mass eigenbasis as follows
\begin{equation}
    \mathcal{L}_Y= Y^u \overline{Q}\tilde{H}_1 u_R + Y^d \overline{Q} H_1 d_R + Y^e \overline{L}H_1 e_R + \rho^u \overline{Q}\tilde{H}_2 u_R + \rho^d \overline{Q} H_2 d_R + \rho^e \overline{L}H_2 e_R + \text{H.c.},
\end{equation}
with $Y^i= \mathrm{diag}(\sqrt{2}M^i)/v$. The $\rho$ matrices can introduce tree-level flavor-changing neutral currents (FCNCs) through nondiagonal entries, and the new scalar-fermion interactions arising from them are completely uncorrelated with fermion masses and mixings. 

In the Higgs basis, $A$ and $H^+$ are the pseudoscalar and charged scalar mass eigenstates with masses
\begin{equation}
    m_A^2=\mu_{22}^2+\frac{(\eta_3+\eta_4-\eta_5)}{2}v^2   \; ,\; m_{H^+}^2=\mu_{22}^2+\frac{\eta_3}{2}v^2.
\end{equation}
By virtue of the stabilizing dark symmetry $Z_2^D$, $S$ is also a mass eigenstate with mass 
\begin{equation}
    m_S^2=\frac{\mu_{S}^2}{2}+\lambda_{11}v^2  \;.
\end{equation}
Only the CP,$Z_2^D$ even neutral scalars need to be diagonalized. We parametrize the $h_1$-$h_2$ mixing as follows 
\begin{equation}
    M_0^2=\begin{pmatrix}
        \eta_1 v^2 & \eta_6 v^2\\
        \eta_6 v^2 & \mu_{22}^2+\frac{(\eta_3+\eta_4+\eta_5)}{2}v^2
    \end{pmatrix}= R \begin{pmatrix}
        m_H^2&0\\
        0&m_h^2
    \end{pmatrix}R^T \;,\; R= \begin{pmatrix}
        c_\gamma & s_\gamma\\
        -s_\gamma & c_\gamma
    \end{pmatrix} ,
\end{equation}
where we have introduced the physical masses $m_h\approx 125$ GeV, $m_H$, and the mixing angle~$\gamma$ with notations $\cos\gamma=c_\gamma$, $\sin\gamma=s_\gamma$. With this definition for the mixing angle, Higgs alignment is achieved in the $c_\gamma\rightarrow0$ limit.

Phenomenologically, the parameters $\gamma$, $\rho^e$, $\rho^u$, $\rho^d$ and the exotic scalar masses are restricted by measurements of the $125$ GeV Higgs properties, direct searches for exotic scalars, exotic meson decays, lepton flavor violation experiments, eEDM and nEDM measurements, and so on. In order for the observed scalar to have SM-like gauge boson couplings, $c_\gamma\sim 0$. 

In this work we examine the possibility that dark matter communicates with the SM through the scalar G2HDM portal, in the $\rho_{tt}$ driven EWBG scenario. According to previous studies, successful EWBG is possible in this scenario provided the scalar potential contains large couplings, which are needed for a strong first-order electroweak phase transition~(FOEWPT). Additionally, a large complex $\rho_{tt}$ can be used to drive~\cite{Fuyuto:2017ewj} CP-violation in scatterings during FOEWPT. If the exotic scalars are in the sub-TeV scale with a relatively large mixing angle $\gamma$, there is a prospect of beyond~SM signals in collider experiments and EDM measurements of the electron and neutron. 
In order to take full advantage of these two ingredients, we focus on the DM phenomenology driven by the $S^2H_1^\dagger H_2$ coupling, which generates an $s$-channel annihilation cross section into top pairs, mediated by $\rho_{tt}$. Due to the doublet nature of the Higgs bosons we must also consider annihilation channels into vector and scalar bosons. We will investigate first the visible scalar sector of the theory to determine the allowed parameter space of the parameters
\begin{equation}
\{\gamma,\rho_{tt},m_H,m_A,m_{H^+}\},
\end{equation}
focusing on the region with large scalar couplings and large, complex $\rho_{tt}$ and with scalar states discoverable at LHC and/or high-luminosity LHC. 

We will focus on the $\lambda_{11}=0=\lambda_{22}$ scenario due to the following considerations. The $\lambda_{11}$ scenario resembles the SM-Higgs mediated singlet dark matter scenario, which is well-known and strongly constrained by direct detection experiments. In this case, departure from alignment will introduce interactions between $S$ and $H$ suppressed by $c_\gamma$. Due to the small value of this parameter allowed by experimental data, these new interactions are expected to be subdominant with respect to $h$-mediated interactions. 
The $\lambda_{22}$ scenario, on the other hand, will involve dark matter interacting directly only with two scalars, which leads to DM annihilating into two scalars at the tree level if kinematically allowed or into SM fermions and vector bosons at one loop if the scalars are too heavy. The first possibility is also included in the $\lambda_{12}$ scenario, so we believe it is not necessary to repeat the analysis. The second possibility could lead to viable phenomenology, but we consider it outside the scope of this paper.
First, we will study the phenomenology of the $Z_2^D$ even scalars, taking into account current constraints on the couplings of the observed Higgs boson and searches for BSM scalars. 
After selecting representative benchmark points for $Z_2^D$ even scalar masses, mixings and $\rho_{tt}$, we proceed to scan over the remaining parameters of the dark sector, $m_S$ and $\lambda_{12}$, to study the parameter space allowed by direct and indirect dark matter detection experiments and relic density constraints. 
%###################################################################
\section{Visible scalar phenomenology}\label{sec:vispheno}

\subsection{Theoretical Constraints}
The scalar potential parameters are required to satisfy both perturbativity and perturbative unitarity constraints.
Parameter points are generated by imposing the tree-level masses of the scalar mass eigenstates, while ensuring that the corresponding quartic couplings satisfy $|\eta_i| < 4\pi$. 
We further verify, using a model file generated by \texttt{SARAH} \cite{Staub:2013tta} in \texttt{SPheno}~\cite{Porod:2003um,Porod:2011nf}, that the perturbative unitarity conditions are satisfied at tree level.

\subsection{Experimental Constraints}
The additional scalar states contribute at the one-loop level to the electroweak gauge-boson vacuum polarization amplitudes, where these effects are parametrized by the oblique parameters $S$, $T$, and $U$~\cite{Peskin:1990zt,Grimus:2007if,Grimus:2008nb}.
These observables provide a sensitive probe of physics beyond the SM through global electroweak precision fits~\cite{ParticleDataGroup:2024cfk}.
In particular, the oblique parameters are sensitive to the mass splittings and mixing among the additional scalar states.
In our analysis, we compute the scalar contributions to $S$, $T$, and $U$ using \texttt{SPheno}, and require them to lie within the experimentally allowed region at the 95\% confidence level~(CL).

\begin{figure}[t]
    \centering
    \includegraphics[width=0.5\linewidth]{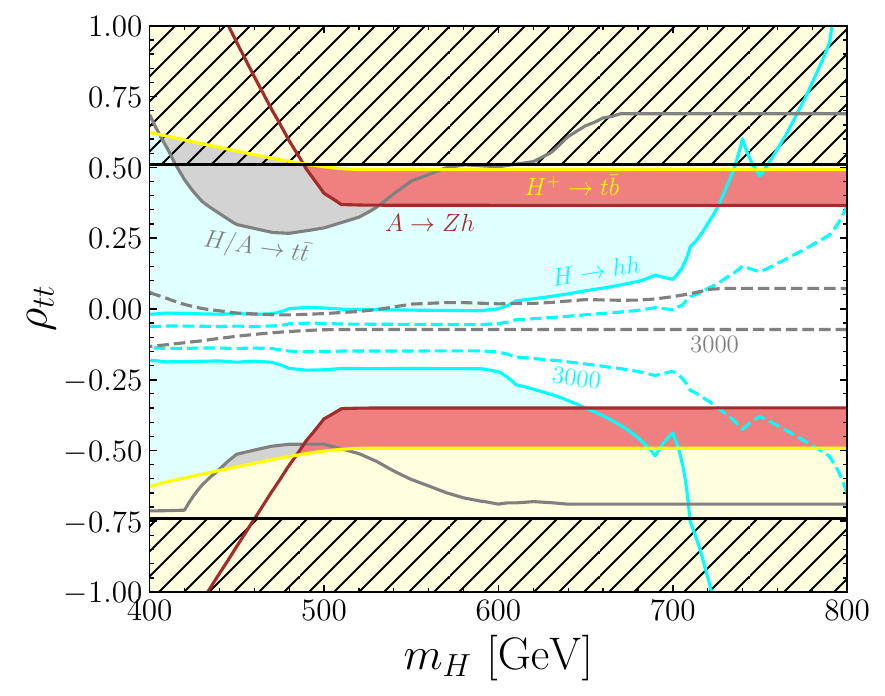}    
    \caption{Excluded $m_H$--$\rho_{tt}$ parameter space at the $95\%$ CL by direct searches for additional Higgs bosons for $c_\gamma = 0.1$, along with the projected HL-LHC reach. The hatched region is excluded by \texttt{HiggsSignals} for $c_\gamma = 0.1$. The Higgs masses are fixed to $m_A = m_{H^+}=600$~GeV.}
    \label{fig:HSHB}
\end{figure}

The observed 125~GeV resonance by ATLAS and CMS experiments at the LHC, which is compatible with the SM Higgs boson, puts severe constraints on the mixing angle $\gamma$, $\rho_{tt}$, and $\eta_i$.
Here we identify the light CP-even scalar $h$ as the observed 125~GeV SM-like Higgs.
The angle $\gamma$ tends to reduce all couplings of $h$ to SM fermions and gauge bosons, while $\rho_{tt}$ in the presence of nonzero $c_\gamma$ modifies the $h$-top quark coupling, and subsequently the one-loop gluon and photon couplings. The values of $\eta_3$ and $\eta_7$ control the $h$ coupling to the charged Higgs $H^+$, which modifies the Higgs decay rate into photons at the one-loop level as well. We use the \texttt{HiggsSignals}~\cite{Bechtle:2013xfa,Bechtle:2020uwn} module of \texttt{HiggsTools}~\cite{Bahl:2022igd} to check that the properties of $h$ match the observed signal rates of $h_{125}$. 
For $c_\gamma = 0.1$, we show in Fig.~\ref{fig:HSHB} the $m_H$--$\rho_{tt}$ parameter space excluded at the $95\%$ CL by \texttt{HiggsSignals} (hatched region). The bounds vanish in the limit of $c_\gamma = 0$.
The \texttt{HiggsSignals} allowed regions in the $c_\gamma$--$\rho_{tt}$ plane and $\mathrm{Re}(\rho_{tt})$--$\mathrm{Im}(\rho_{tt})$ plane are plotted in Fig~\ref{appfig:HS} of appendix~\ref{appendix:HS}. 

The presence of sub-TeV exotic scalars, $H$, $A$ and $H^+$, in the visible sector opens up the possibility of probing the model at the LHC. The scalar mixing $c_\gamma$ introduces SM Higgs-like couplings to $H$ and $\rho_{tt}$ couples $H$ and $A$ to the top, and $H^+$ to $\overline{b}t$, in addition to the electroweak couplings such as $HWW$, $HZZ$, $HZA$, $hZA$, $HW^-H^+$ and $hW^-H^+$, which are controlled by the mixing angle $\gamma$.
We examine current LHC constraints with the help of \texttt{HiggsBounds}~\cite{Bechtle:2008jh,Bahl:2021yhk} module of \texttt{HiggsTools}. We fix $m_A=m_{H^+}=600$ GeV, $\eta_{2,7} = 0.01$, $\mu_{22}^2=(600\text{ GeV})^2$, $c_\gamma=0.1$, and scan over the ranges $400\leq m_H\leq800~\text{ GeV}$, and $-1\leq \rho_{tt}\leq1$. 
In Fig.~\ref{fig:HSHB}, we show the exclusion bounds at 95\% CL in the $m_H$--$\rho_{tt}$ plane, where the most sensitive limits come from searches for $H \to hh$~\cite{ATLAS:2023vdy,ATLAS:2022xzm,CMS:2024phk}, $H/A \to t\bar t$~\cite{CMS:2025dzq},\footnote{The limits are applied by hand, and omit toponium contribution. However, including the toponium contribution would mildly relax the bounds.} $A \to Zh$~\cite{ATLAS:2022enb}, and $H^+ \to t\bar b$~\cite{ATLAS:2021upq}.
The naive expected reach of the HL-LHC is also shown in Fig.~\ref{fig:HSHB}.
We point out that for the $c_\gamma=0$ case, the leading limits would be simply $H/A \to t\bar t$~\cite{CMS:2025dzq} and $H^+ \to t\bar b$~\cite{ATLAS:2021upq}.
We note that turning on the flavor-violating $\rho_{tc}$ coupling would induce Higgs decays such as $H/A \to t\bar c$ and $H^+ \to c\bar b$ and thus soften the exclusion bounds. 
Further constraints on $\rho_{tt}$ and Higgs masses from direct searches for G2HDM scalars can be found in Refs.~\cite{ATLAS:2023tlp,CMS:2023xpx,CMS:2025plw}.

%###################################################################
\section{Dark Matter Phenomenology}\label{sec:DMpheno}
\subsection{Relic Density}
\begin{figure}[t]
    \centering
    \includegraphics[width=0.9\linewidth]{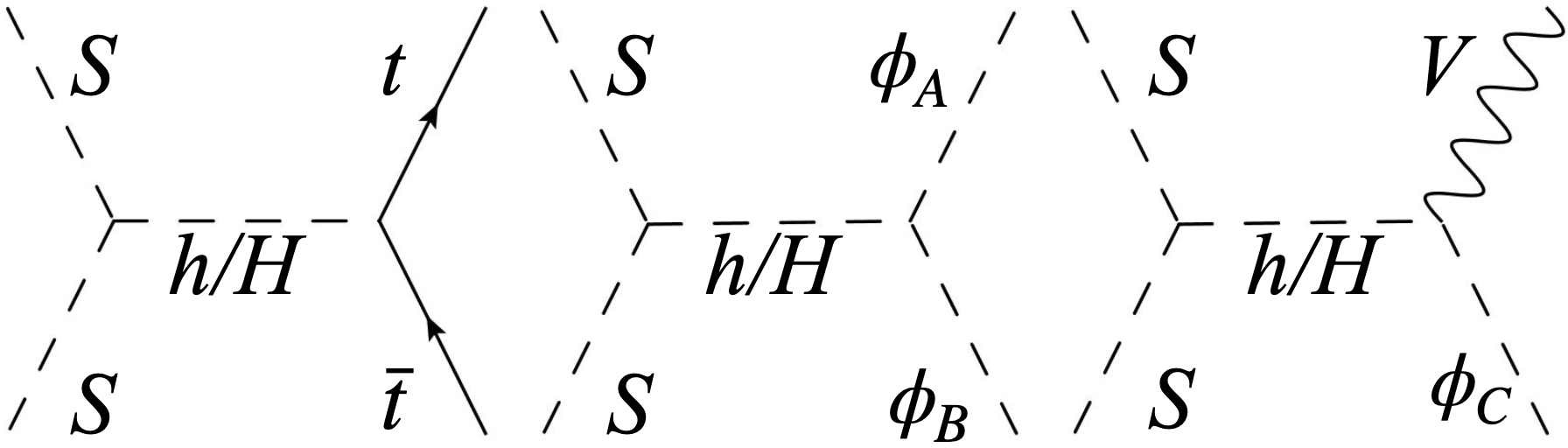}
    \caption{Leading contributions to dark matter freeze-out annihilation cross section. Here the possible final states in the second diagram are $\{\phi_A,\phi_B \}=\{h,h\},\{h,H\},\{H,H\},\{A,A\}$ or $\{H^+,H^-\}$, and in the third diagram are $\{V,\phi_C \}=\{Z,A\}$ or $\{W^\pm,H^\mp \}$.}
    \label{fig:relicdiagrams}
\end{figure}

This model contains a scalar portal to the dark sector mediated by the two CP-even scalars $h$ and $H$. Due to the CP invariance of the scalar potential, the $\lambda_{12}$ term does not contain an $SSA$ coupling.
The relic density of the dark matter candidate $S$ in the parameter space defined previously, for masses $m_S\geq m_t$, is determined at the tree level by the diagrams in Fig.~\ref{fig:relicdiagrams}. 
We calculate the relic density from thermal freezeout with the \texttt{micrOMEGAs} code~\cite{Alguero:2023zol}. By setting $m_H<m_A,m_{H^+}$ the first diagram dominates in the mass range $m_t<m_S<m_H$, encountering a resonant enhancement at $2m_S\sim m_H$. As $m_S$ rises and reaches $m_S\sim m_A,m_{H^+}$, the second and third diagrams become kinematically accessible and dominate the annihilation cross section. For masses $m_S\leq m_t$, the leading contribution to dark matter annihilation are the $h$-mediated channels into SM fermions. In this region, a resonant enhancement at $2m_S\sim m_h$ is also present.
\subsection{Direct Detection}
\begin{figure}
    \centering
    \includegraphics[width=0.5\linewidth]{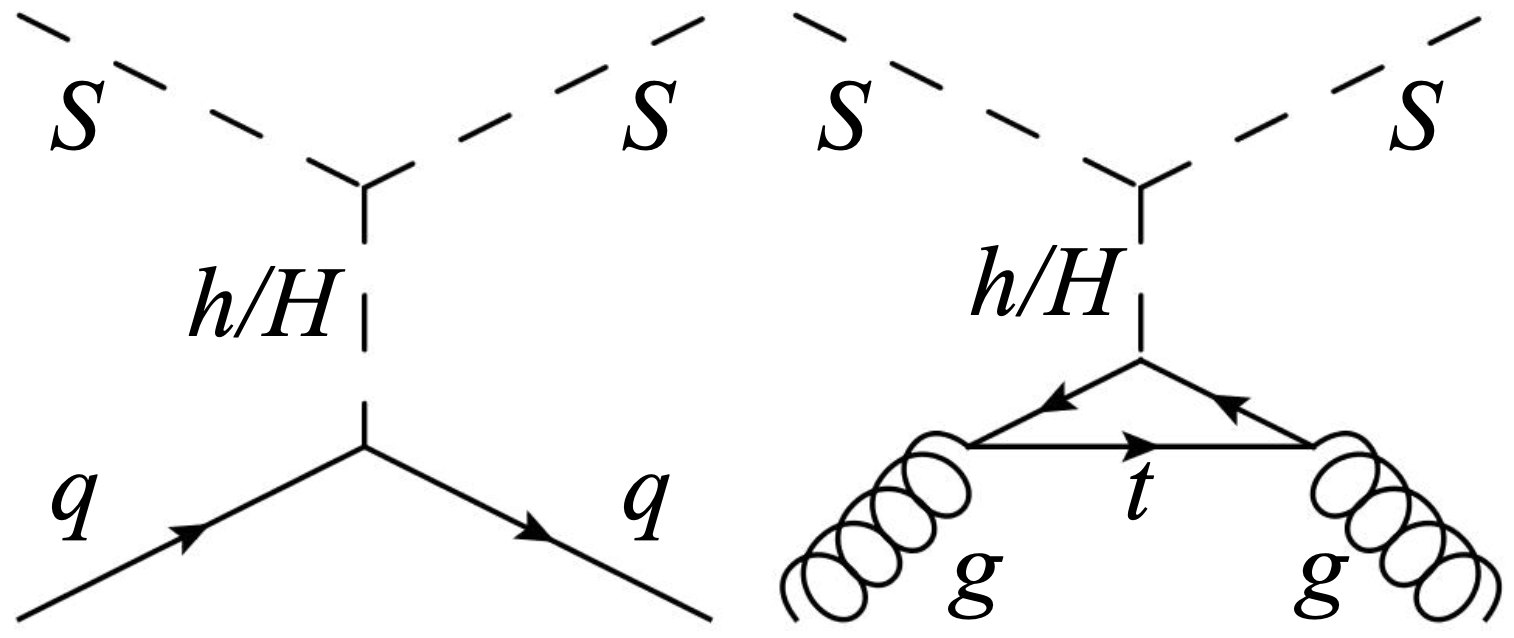}
    \caption{Leading contributions to the elastic dark matter-nucleon scattering cross section.}
    \label{fig:directdetection}
\end{figure}
The leading contributions to the direct detection of dark matter in elastic nucleon scattering experiments are shown in Fig.~\ref{fig:directdetection}. These contributions induce the effective DM-parton operators $\mathcal{O}_{3,q}$ and  $\mathcal{O}_6$ of \cite{Bishara:2016hek}. These are
\begin{equation}
\mathcal{O}_{3,q}= m_q(SS)(\bar{q}q) \; ;\;\mathcal{O}_{6}= \frac{\alpha_s}{12\pi}(SS)G_{\mu\nu}G^{\mu\nu},
\end{equation}
where $q=u,d,s$ are the light quarks and $G_{\mu\nu}$ is the gluon strength tensor. The corresponding Wilson coefficients induced by the diagrams in Fig. \ref{fig:directdetection} are
\begin{eqnarray}
    \mathcal{C}_6&=&\frac{12 \pi\lambda_{12}v}{\alpha_s}\left( \frac{s_\gamma g^H_{gg}}{m_H^2}+ \frac{c_\gamma g^h_{gg}}{m_h^2}\right),\\
    \mathcal{C}_{3,q}&=&\sqrt{2}\lambda_{12}s_\gamma c_\gamma\left( \frac{1}{m_H^2}- \frac{1}{m_h^2}\right),
\end{eqnarray}
where $g^{H(h)}_{gg}$ is the gluon coupling to $H$($h$). These effective operators lead to nucleon-dark matter elastic scattering. Using heavy baryon chiral perturbation theory \cite{Bishara:2016hek}, the spin-dependent nuclear differential cross section is 
\begin{equation}
    \frac{d\sigma}{dE_R}=\frac{2m_A}{(2J_A+1) |v_S|^2}\left(\sum_{\tau,\tau '}R^{\tau\tau '}_M W^{\tau\tau '}_M(q)\right),
\end{equation}
where $\tau,\tau '=n,p$ are nucleon isospin indices, $W^{\tau\tau´}_M(q)$ is the nuclear response function, and $R^{\tau\tau´}_M(q)$  is given by 
\begin{eqnarray}
    R^{\tau\tau´}_M(q)&=&c^{(0)}_{1,\tau}c^{(0)}_{1,\tau '}\\
    c^{(0)}_{1,\tau}c^{(0)}&=&\frac{-2 m_G}{27}\mathcal{C}_6+\sigma^\tau_u\mathcal{C}_{3,u}+\sigma^\tau_d\mathcal{C}_{3,d}+\sigma^\tau_s\mathcal{C}_{3,s}.
\end{eqnarray}
Here, $m_G$ is the gluon contribution to the nuclear mass and $\sigma_{u,d}^\tau$ are the axial vector matrix elements \cite{Bishara:2016hek}. 
We compare the obtained elastic dark matter-nucleon spin-independent cross section to the limits set by the LUX-ZEPLIN~(LZ) collaboration \cite{LZ:2024zvo}.

\begin{figure}[t]
    \centering
\includegraphics[width=0.45\linewidth]{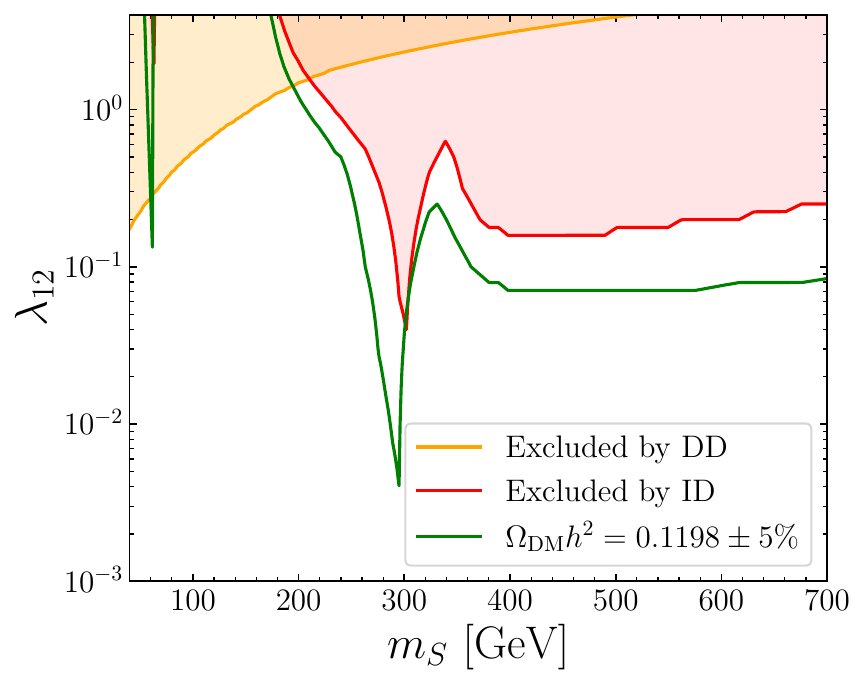}
\includegraphics[width=0.45\linewidth]{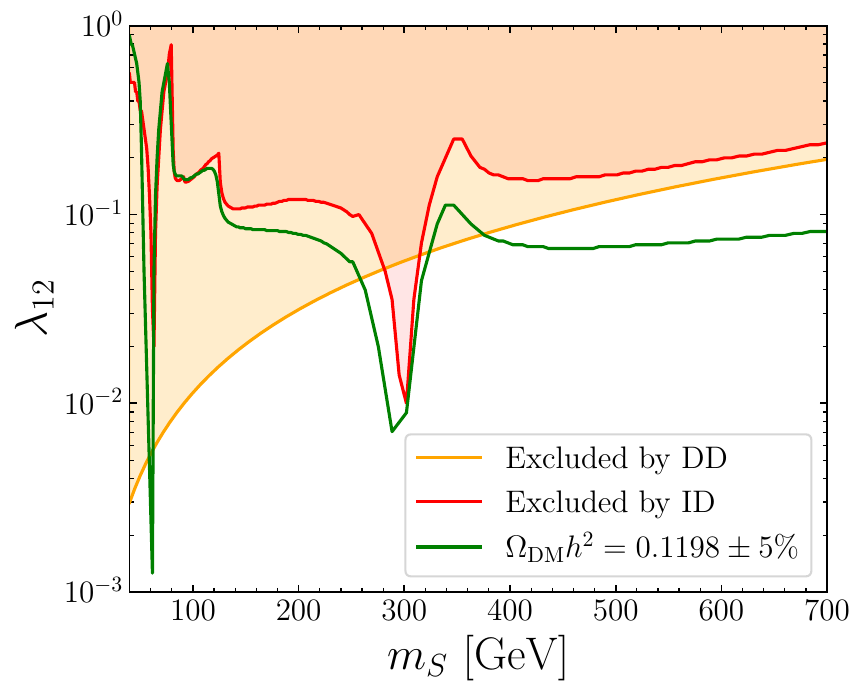}
    \caption{Excluded $m_S$--$\lambda_{12}$ parameter space by direct detection LZ results (orange) and indirect detection limits of positrons and protons (red). The correct relic density is obtained at the green line. Here $m_H = m_A = m_{H^+}=600$ GeV, $|\rho_{tt}|=0.1$, $c_\gamma=0.001$~(left), and $c_\gamma = 0.1$~(right).} 
    \label{fig:dmscan}
\end{figure}

\subsection{Indirect Detection}
The annihilation of dark matter into SM particles in galactic haloes can produce an excess in the spectrum of gamma rays, positrons or antiprotons over the expected background from SM astrophysical processes.  We calculate the annihilation cross section of dark matter particles in haloes with \texttt{micrOMEGAs}, comparing the velocity-averaged annihilation cross section into top pairs and gauge bosons with the stringent limits from the measurement of cosmic antiproton flux by AMS-02 \cite{Cuoco:2017iax} and into bottom quark pairs from the measurement of the gamma ray spectrum of dwarf spheroidal galaxies by Fermi-LAT \cite{Fermi-LAT:2015att}. The dark matter annihilation cross section proceeds in the same channels as for the determination of dark matter relic density. As such, this cross section also presents resonant regions near $2m_S \sim m_h,m_H$. 
\subsection{Dark Matter Numerical Scan}
We perform a numerical scan for two scenarios, chosen as to be compatible with the previously discussed constraints for $h, H, A$ and $H^+$ at the LHC. The chosen parameters for the scan\footnote{Throughout, we assume $\rho_{tt}$ to be real, since a nonzero complex phase would not affect dark matter phenomenology (see appendix~\ref{appendix:phitt}). Such a phase is strongly constrained by electron EDM bounds; however, a cancellation mechanism~\cite{Fuyuto:2019svr} allows one to survive these constraints.} are \begin{eqnarray}
    m_H=m_A&=&m_{H^+}=600~\text{GeV} \nonumber\\
    c_\gamma&=&0.001,0.1\nonumber\\
    \eta_{2,7}&=&0.01\nonumber\\
    \mu^2_{22}&=&(600~\text{GeV})^2\nonumber\\
    |\rho_{tt}|&=&0.1\nonumber\\
    \lambda_{12}&\in& (10^{-3},10)\nonumber\\
     m_S&\in& (50,700)\text{ GeV.}
\end{eqnarray}
The results of the numerical scan are shown in Fig.~\ref{fig:dmscan}. In both panels of this figure, we show the regions excluded by direct and indirect searches for DM, as well as the region consistent with the observed relic density. In the left panel we depict the scan for $c_\gamma=0.001$, showing that at a smaller mixing angle the direct detection cross section is suppressed. Both the relic density and indirect detection constraints show resonance peaks owing to the same $s$-channel contributions to them. The combined direct detection and relic density constraints exclude DM masses below 200 GeV, except for a narrow region near the $h$ mass $s$-channel resonance ($m_S\approx m_h/2$). We also note that indirect detection constraints are close to the curve where relic density is satisfied, but not yet excluding it. We observe that as the mass of the DM candidate grows, the indirect detection bound relaxes more quickly than the relic density constraint.
In the right panel of Fig. \ref{fig:dmscan}, we show the scan results for $c_\gamma=0.1$. We observe that a higher value of the scalar mixing angle enhances the direct detection cross section. Combined with the relic density constraint, this excludes DM masses below 380 GeV, with the exception of the narrow $m_S\approx m_h/2$ resonance and the wider $m_S\approx m_H/2$ resonance.

%###################################################################
\section{Benchmark Scenarios and Collider Signatures}\label{sec:BPs}
%\subsection{Benchmark Points}
\begin{table}[t]
    \centering
    \setlength{\tabcolsep}{10pt}
\begin{tabular}{|c|c|c|c|c|c||c|}
       \hline
       & BP1 & BP2 & BP3 & BP4 & BP5 & BP1$^\prime$ \\ \hline
        %$\rho_{tc}$  & 0 & 0 &0 &0 &0 & $10^{-1}$ \\ \hline
        $m_{H}$ & 450 & 800 & 600 & 600 & 600 & 450 \\ \hline
        $m_{A}$ & 600 & 600 & 400 & 800 & 600 & 600 \\ \hline
        $m_{H^+}$ & 600 & 600 & 600 & 600 & 600 & 600 \\ \hline
        $m_{S}$ & 180 & 300 & 250 & 250 & 230 & 180 \\ \hline
        %$c_\gamma$&$10^{-3}$  &$10^{-3}$  & $10^{-3}$&$10^{-3}$ &$10^{-3}$ &$10^{-3}$ \\ \hline
       $\lambda_{12}$ & 0.75 & 0.1 & 0.15 & 0.34 & 0.6 & 0.2 \\ \hline\hline
       $\sigma_{\rm{DD}}/10^{-12}~[\rm{pb}]$ & 4.6 & 0.007 & 0.05 & 0.25 & 0.94 & 2.1 \\ \hline
       $\mathcal{B}(H\rightarrow SS)$ & 0.95 & 0.0005 & 0.006 & 0.54 & 0.72 & 0.30 \\ \hline
       $\mathcal{B}(H\rightarrow ZA)$ & -- & 0.3 & 0.97 & -- & -- & -- \\ \hline
       $\mathcal{B}(H\rightarrow W^- H^+)$ & -- & 0.68 & -- & -- & -- & -- \\ \hline
       $\mathcal{B}(H\rightarrow t\bar{t})$ & 0.046 & 0.01 & 0.025 & 0.45 & 0.27 & 0.18 \\ \hline
       $\mathcal{B}(H\rightarrow t\bar{c})$ & 0 & 0 & 0 & 0 & 0 & 0.51 \\ \hline
       $\mathcal{B}(A\rightarrow ZH)$ & 0.90 & -- & -- & 0.31 & -- & 0.82 \\ \hline
       $\mathcal{B}(A\rightarrow W^- H^+)$ & -- & -- & -- & 0.67 & -- & 0 \\ \hline
       $\mathcal{B}(A\rightarrow t\bar{t})$ & 0.096 & 1 & 1 & 0.01 & 1 & 0.09 \\ \hline
       $\mathcal{B}(A\rightarrow t\bar{c})$ & 0 & 0 & 0 & 0 & 0 & 0.09 \\ \hline
       $\mathcal{B}(H^+\rightarrow W^+ H)$ & 0.89 & -- & -- & -- & -- & 0.82 \\ \hline
       $\mathcal{B}(H^+\rightarrow W^+ A)$ & -- & -- & 0.95 & -- & -- & -- \\ \hline
       $\mathcal{B}(H^+\rightarrow t\bar b)$ & 0.11 & 1 & 0.05 & 1 & 1 & 0.1 \\ \hline
       $\mathcal{B}(H^+\rightarrow c\bar b)$ & 0 & 0 & 0 & 0 & 0 & 0.07 \\ \hline
\end{tabular}
    \caption{Characteristics of the proposed benchmark points for future collider experiments. For all benchmark points, $|\rho_{tt}| = 0.1$, $\rho_{tc} = 0$, and $c_\gamma = 0.001$, except for BP1$^\prime$ where $\rho_{tc}$ is turned on and fixed to $\rho_{tc} = 0.1$. All masses are in GeV.}
    \label{tab:BPs}
\end{table}

To illustrate the range of phenomenological behaviour realized within the viable
parameter space, Table~\ref{tab:BPs} presents six representative benchmark points, BP1, BP2, BP3, BP4, BP5 and BP1$^\prime$, each chosen to highlight a qualitatively different decay topology among the heavy G2HDM scalars $H$, $A$, $H^\pm$ and the DM state $S$. BP1$^\prime$ shares the mass spectrum of BP1 but switches on a nonzero flavor-violating top-charm coupling (along with a smaller
portal coupling $\lambda_{12}$), and is included specifically to illustrate the effect of non-diagonal Yukawa couplings on the phenomenology. 
For all benchmark points dark matter relic density is within experimental range.

\textbf{BP1} ($m_H=450$~GeV, $m_A=m_{H^+}=600$~GeV, $m_S=180$~GeV, $\lambda_{12}=0.75$)
realizes the singlet-portal regime. The light singlet and the large portal coupling
$\lambda_{12}$ together open and enhance $H\to SS$, which saturates the $H$ width
($\mathcal{B}=0.95$) at the expense of $H\to t\bar t$ (4.6\%); $H\to ZA$ and
$H\to W^-H^+$ are kinematically closed since $m_A=m_{H^+}>m_H$. The same large
$\lambda_{12}$ drives the largest spin-independent direct-detection cross section
of the four points, $\sigma_{\rm DD}=4.6\times 10^{-12}$~pb. With
$m_A=m_{H^+}=600$~GeV both lying above $m_H+m_Z$, the pseudoscalar decays
dominantly through $A\to ZH$ (90\%), with $A\to t\bar t$ a small remainder; the
charged Higgs, degenerate with $A$, decays almost entirely through
$H^+\to W^+H$ (89\%), with $H^+\to t\bar b$ making up the rest. BP1 therefore
probes a scenario in which Higgs-to-Higgs cascades ($A\to ZH$, $H^+\to W^+H$,
$H\to SS$) dominate over the conventional fermionic final states, while
simultaneously yielding the strongest DM direct-detection signal among the
benchmarks.

\textbf{BP2} ($m_H=800$~GeV, $m_A=m_{H^+}=600$~GeV, $m_S=300$~GeV, $\lambda_{12}=0.1$)
instead realizes a regime with a heavier, more weakly-coupled singlet,
suppressing $H\to SS$ to a negligible $5\times10^{-4}$ and reducing
$\sigma_{\rm DD}$ to $7\times10^{-15}$~pb, the smallest value in the table. With
$m_H=800$~GeV well above $m_A=m_{H^+}=600$~GeV, the bosonic channels
$H\to ZA$ (30\%) and $H\to W^-H^+$ (68\%) dominate the $H$ width, leaving only a 1\% $t\bar t$ fraction. Because $A$ and $H^+$ are degenerate, the bosonic channels $A\to ZH$ and $H^+\to W^+H$ are kinematically forbidden, and both instead decay exclusively to third-generation quarks:
$\mathcal{B}(A\to t\bar t)=\mathcal{B}(H^+\to t\bar b)=1$. BP2 thus illustrates a scenario dominated by $H\to W^-H^+, ZA$ bosonic channels alongside conventional fermionic decays of the mass-degenerate $A$ and $H^+$, with essentially no DM direct-detection sensitivity.

\textbf{BP3} ($m_H=m_{H^+}=600$~GeV, $m_A=400$~GeV, $m_S=250$~GeV,
$\lambda_{12}=0.15$) places the pseudoscalar well below $H$ and $H^+$, which remain degenerate at 600~GeV. Since $m_A<m_H$, the channel $H\to ZA$ is  kinematically open, and indeed dominates the $H$ width (97\%); the small $\lambda_{12}$ together with a relatively light singlet leaves $H\to SS$ strongly suppressed (0.6\%), and $H\to t\bar t$ is reduced to a 2.5\% remainder, with a correspondingly small $\sigma_{\rm DD}=0.05\times10^{-12}$~pb. Because $H^+$ is degenerate with $H$ rather than with $A$, $H^+\to W^+H$ is closed, while $H^+\to W^+A$ is open by the full 200~GeV mass gap and dominates the charged Higgs width (95\%), with only a 5\% $H^+\to t\bar b$ remainder. The pseudoscalar, being the lightest of the three, has no bosonic channel available to it and decays essentially entirely to $t\bar t$ ($\mathcal{B}=1$). BP3 therefore represents a benchmark in which a single bosonic cascade, $H\to ZA$, feeds the $H$ width while $H^+\to W^+A$ provides a second, independent bosonic channel for the charged Higgs, with $A$ itself decaying purely fermionically.
 
\textbf{BP4} ($m_H=m_{H^+}=600$~GeV, $m_A=800$~GeV, $m_S=250$~GeV, $\lambda_{12}=0.34$) flips the pseudoscalar from the lightest state in BP3 to the heaviest here, while $H$ and $H^+$ remain degenerate at 600~GeV in both benchmarks. With $m_A=800$~GeV now above $H$ and $H^+$, $H\to ZA$ is closed, and instead the larger portal coupling lets $H\to SS$ (54\%) and $H\to t\bar
t$ (45\%) split the $H$ width comparably, giving $\sigma_{\rm DD}=0.25\times10^{-12}$~pb. The pseudoscalar, now 200~GeV above $H$ and $H^+$, decays through both bosonic channels open to it, $A\to W^-H^+$ (67\%) and $A\to ZH$ (31\%), with only a 1\% $t\bar t$ fraction. The charged Higgs, degenerate with $H$, decays exclusively to $H^+\to t\bar b$
since $H^+\to W^+H$ and $H^+\to W^+A$ are both kinematically closed, and $H^+ \to W^+ h$ is suppressed by $c_\gamma$.
BP4 thus combines a large $H\to SS$ branching ratio with $A$ decays dominated by bosonic channels rather than $t\bar t$, while the charged Higgs decays purely fermionically, offering a complementary search target to BP1--BP3.

\textbf{BP5} ($m_H=m_A=m_{H^+}=600$~GeV, $m_S=230$~GeV, $\lambda_{12}=0.6$)
may appear at first sight similar to BP1, since $H\to SS$ again dominates the
$H$ width (72\%), but the two benchmarks are physically quite distinct. In BP1, the heavier $A$ and $H^+$ could decay to $H$ via $A\to ZH$ (90\%) and $H^+\to W^+H$ (89\%), so that a large fraction of the $A$ and $H^+$ production cross section ultimately feeds into $H\to SS$, enhancing the missing-energy signal beyond the direct $gg\to H$ contribution alone. In BP5,
by contrast, the full degeneracy $m_H=m_A=m_{H^+}=600$~GeV closes every bosonic decay channel, leaving $A$ and $H^+$ to decay purely
fermionically: $\mathcal{B}(A\to t\bar t)=\mathcal{B}(H^+\to t\bar b)=1$. The
$H\to SS$ signal therefore receives no cascade feeding from $A$ or $H^+$ and
must rely entirely on direct $H$ production. The larger portal coupling
$\lambda_{12}=0.6$, together with $2m_S=460~\text{GeV}<m_H$, keeps $H\to SS$
dominant and yields $\sigma_{\rm DD}=0.94\times10^{-12}$~pb, roughly five times
smaller than BP1. BP5 thus represents the
starkest test of the singlet-portal mechanism in isolation: any missing-energy
excess must come from direct $H$ production alone, while the degenerate $A$ and
$H^+$ yield only $t\bar t$ and $t\bar b$ final states that are
indistinguishable from SM top backgrounds, making the DM signal extraction
considerably more challenging than in BP1.

\textbf{BP1$'$} shares the same mass spectrum as BP1
($m_H=450$~GeV, $m_A=m_{H^+}=600$~GeV, $m_S=180$~GeV) but with a somewhat smaller portal coupling, $\lambda_{12}=0.2$, and a non-zero flavor-violating top-charm coupling $\rho_{tc}$. It is included specifically to isolate the effect of non-diagonal Yukawa couplings on the phenomenology. Turning on the $\rho_{tc}$ coupling induces channels that are absent in BP1: $\mathcal{B}(H\to t\bar c)=0.51$, $\mathcal{B}(A\to t\bar c)=0.09$ and $\mathcal{B}(H^+\to c\bar b)=0.07$. For $H$, the flavor-violating channel becomes competitive with $H\to SS$ (30\%) and even exceeds the flavor-conserving $H\to t\bar t$ (18\%), so that $H\to t\bar c$ is in fact the single largest decay mode of $H$ in this benchmark. For $A$, by contrast, the bosonic cascade $A\to ZH$ remains the dominant channel by far, though its branching ratio drops somewhat relative to BP1 (90\%$\to$82\%) as the $t\bar t$ channel is partially diluted into the newly opened $t\bar c$ mode; $A\to t\bar t$ and $A\to t\bar c$ end up sharing the small remainder almost equally (9\% each). The charged Higgs behaves similarly: $H^+\to W^+H$ remains the dominant channel (82\%, down from 89\% in BP1), while $H^+\to t\bar b$~(10\%) and $H^+\to c\bar b$ (7\%) split the remainder. 

Comparing BP1 and BP1$'$ thus shows that, once a flavor-violating $\rho_{tc}$ coupling is present, it can compete on equal footing with the singlet-portal and flavor-conserving fermionic channels in determining the $H$ width, to the point of becoming the predominant $H$ decay mode. The bosonic decay of $A$ and $H^+$, are structurally unaffected by $\rho_{tc}$; their modest reduction in branching ratio relative to BP1 ($A\to ZH$: 90\%$\to$82\%; $H^+\to W^+H$: 89\%$\to$82\%) is due to the new $tc$ channel opening up and sharing the total width, not a dynamical suppression. Likewise, $\sigma_{\rm DD}$ is governed by the mixing and the portal coupling $\lambda_{12}$ and is {\it insensitive} to $\rho_{tc}$, lowering to $2.1\times10^{-12}$~pb due to the reduction of the portal coupling with respect to BP1. 
It should be noted that the reduction of $\mathcal{B}(H\to SS)$ from 95\% in BP1 to 30\% in BP1$'$ is not a consequence of the $tc$ coupling alone, but reflects also the smaller portal coupling $\lambda_{12}$ (0.75$\to$0.2) chosen for this point; the two effects are therefore not directly comparable without fixing $\lambda_{12}$. This makes BP1$'$ a useful benchmark for collider searches targeting flavor-violating final states such as $H \to t\bar c$, $A \to ZH \to Zt\bar c$ and $H^+\to W^+H \to W^+ t\bar c$, which are more visible and complementary to $H\to SS$ signatures already present in BP1.

\begin{figure}[t]
    \centering
    \includegraphics[width=0.45\linewidth]{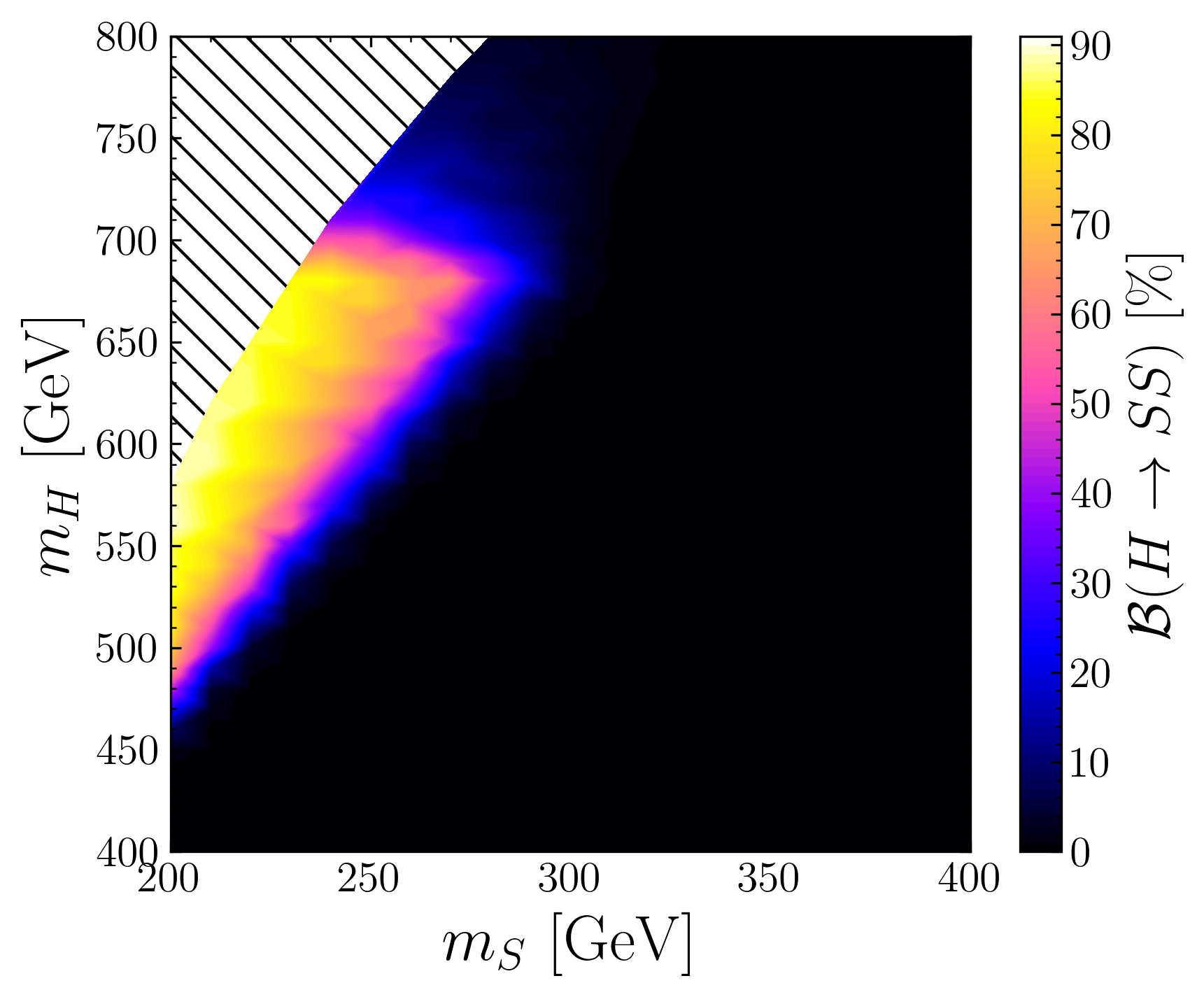}
    \includegraphics[width=0.45\linewidth]{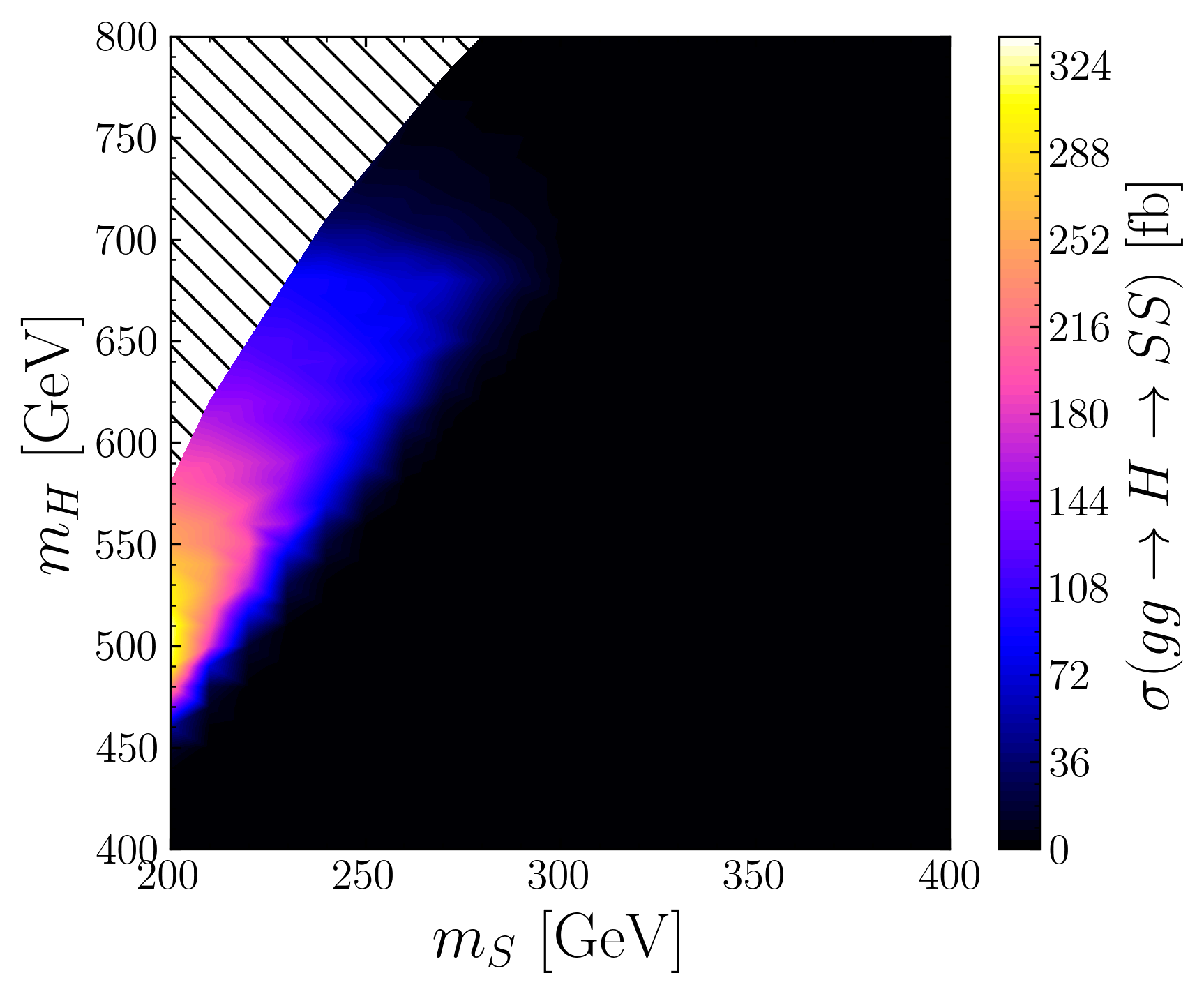}
    \includegraphics[width=0.45\linewidth]{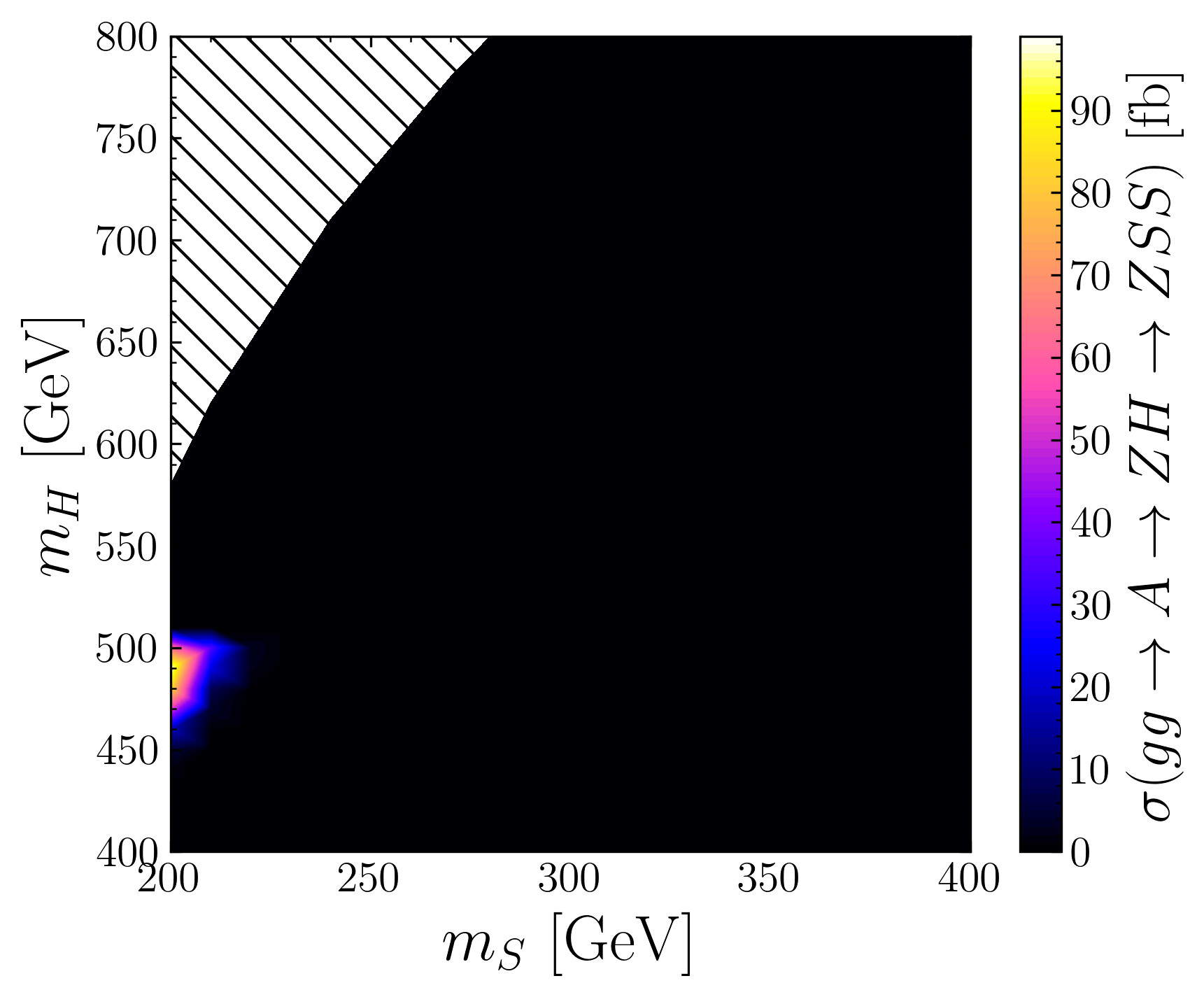}
    \includegraphics[width=0.45\linewidth]{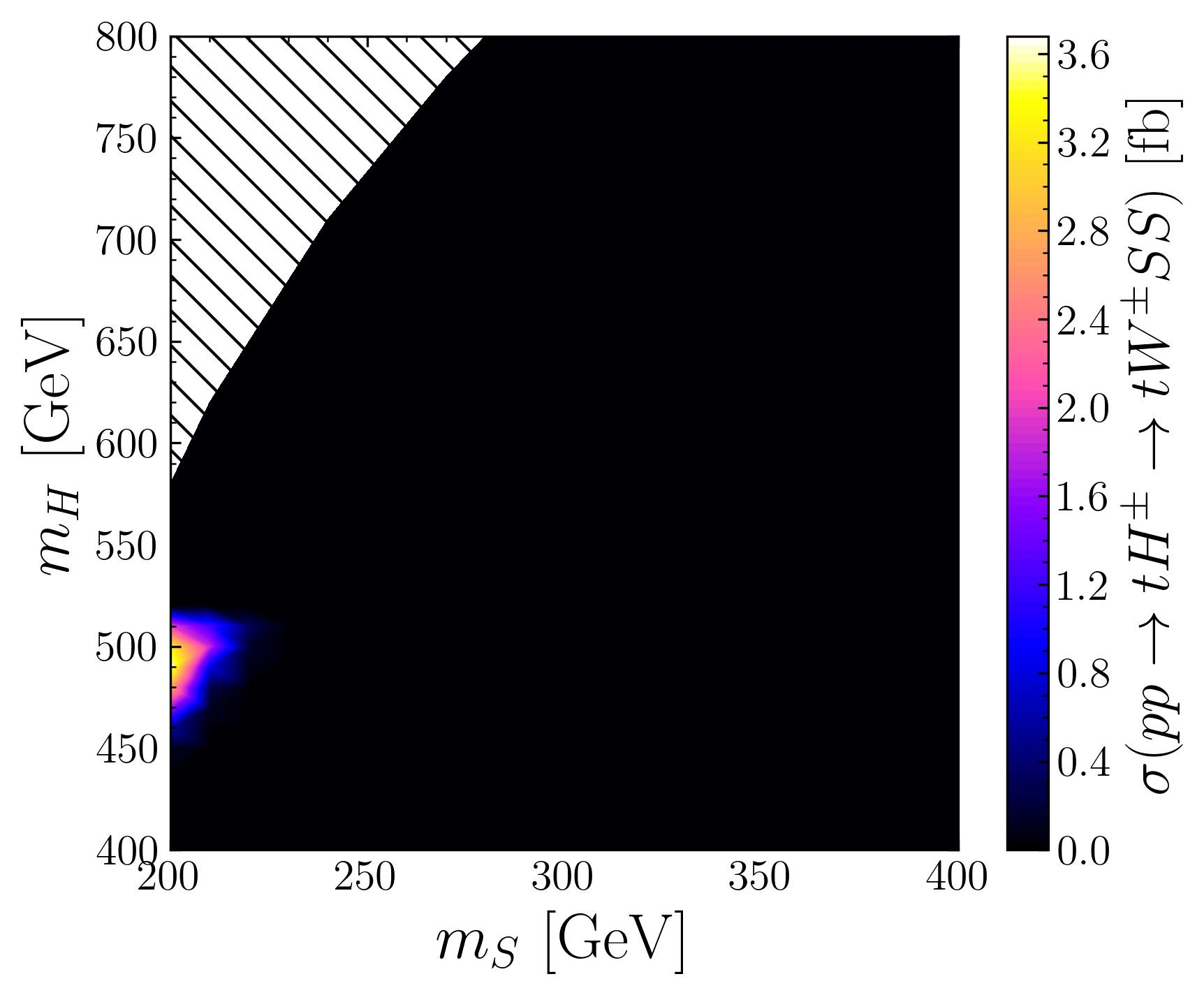}    
    \caption{The branching ratio $\mathcal{B}(H\to SS)$~(top left), the cross sections $\sigma(gg\to H\to SS)$~(top right), $\sigma(gg\to A\to ZH\to ZSS)$~(bottom left), and $\sigma(pp\to tH^{\pm}\to tW^{\pm}H\to tW^{\pm}SS)$~(bottom right) in the $m_S$--$m_H$ plane.
    The hatched region is excluded by direct detection constraints. The production cross sections are at $\sqrt{s} = 14$~TeV.} 
    \label{fig:signatures}
\end{figure}

Beyond these decay-level effects, the non-zero $\rho_{tc}$ coupling also induces production modes that are absent in BP1: single-top associated production $cg \to tH$ and $cg \to tA$, as well as $V_{tb}$-enhanced single charged-Higgs production $cg \to bH^+$. These channels provide production handles that greatly complement the ones already present in BP1, since they tag the heavy scalars with energetic top and bottom quarks in the final state. In particular, the associated production $cg \to tH$ followed by $H\to SS$ allows the top quark to serve as a trigger/tag for an otherwise largely invisible final state, directly probing the dark-matter candidate $S$ through a top quark and missing transverse energy ({\it mono-top} signature). 
This constitutes a novel realization of flavor-violating DM production at the LHC~\cite{Kamenik:2011nb}, distinct from existing simplified model interpretations, in which the heavy Higgs boson acts as the portal between the top sector and the dark matter candidate.
This is a characteristic feature of the G2HDM+S: the non-diagonal Yukawa 
structure, absent in 2HDM+S models with NFC, opens flavor-violating DM production channels that serve as a distinctive signature of the general Yukawa structure of the G2HDM.
Importantly, the {\it mono-top} signature arising from $cg\to tH\to tSS$ would populate the signal regions of existing ATLAS top quark and missing transverse energy searches~\cite{ATLAS:2024xne}, which can constrain the G2HDM+S parameter space, and in particular BP1$'$ scenario. A reinterpretation of existing ATLAS data~\cite{ATLAS:2024xne} in terms of the G2HDM+S, together with a full signal-to-background analysis including detector simulation, is left for future work.

Having identified representative benchmark points that capture distinct phenomenological regimes of the G2HDM+S, we now turn to a broader mapping of the collider signatures across the parameter space. Rather than focusing on isolated points, we perform a two-dimensional scan over the singlet mass $m_S$ and the heavy CP-even Higgs mass $m_H$, and show, in Fig.~\ref{fig:signatures}, the branching ratio $\mathcal{B}(H\to SS)$, which governs the portal activity of $H$, and the cross section $\sigma(gg\to H\to SS)$, which reflects the DM signal yield from direct $H$ production. We further show two complementary cascade channels: $\sigma(gg\to A\to ZH\to ZSS)$, which requires $m_A > m_H + m_Z$ and $m_H > 2m_S$ and provides a $Z$-tagged handle on the same DM final state; and $\sigma(pp\to tH^{\pm}\to tW^{\pm}H\to tW^{\pm}SS)$, which also requires $m_{H^+} > m_H + m_W$ and $m_H > 2m_S$, yielding a $tW$ plus missing transverse energy final state. 

%###################################################################
\section{Conclusion}\label{sec:concl}
In this work, we have studied the phenomenology of G2HDM extended by an additional real scalar singlet (G2HDM+S), providing a viable candidate of scalar dark matter. We have shown that this framework offers a UV completion of top-window dark matter scenarios, in which the singlet communicates with the SM predominantly through the top quark.
We have systematically analyzed the constraints on the visible scalar sector from Higgs signal strength measurements and direct searches for additional Higgs bosons at the LHC, and those on the dark sector from cosmological observables, including the observed DM relic density and spin-independent direct-detection as well as indirect detection. The combination of these constraints carves out a well-defined viable region of parameter space that nonetheless admits a rich and diverse LHC phenomenology.

Within this viable parameter space, we have proposed five benchmark scenarios (BP1--BP5) together with a flavor-violating variant (BP1$'$), each designed to highlight a qualitatively distinct decay topology. BP1 exemplifies the singlet-portal regime, where $H\to SS$ saturates the $H$ width and Higgs-to-Higgs cascades $A\to ZH$ and $H^+\to W^+H$ further feed the missing transverse energy signal. BP2 illustrates the regime in which the bosonic decays $H\to ZA$ and $H\to W^-H^+$ dominate, and $A$/$H^+$ decay fermionically. BP3 and BP4 explore complementary mass hierarchies in which $H\to ZA$ and $H^+\to W^+A$, or $A\to ZH$ and $A\to W^-H^+$, provide independent bosonic handles depending on whether $A$ is the lightest or heaviest of the heavy scalars. BP5 isolates the singlet-portal mechanism in a fully mass-degenerate spectrum, where all bosonic decays are closed, and the DM signal must be extracted from direct $H$ production alone. Finally, BP1$'$ demonstrates that a non-zero top-charm coupling $\rho_{tc}$ can make $H\to t\bar{c}$ predominant $H$ decay mode while leaving 
DM-related observables untouched, and simultaneously opens the associated production mode $cg \to tH$ that allows the top quark to tag an otherwise invisible $H\to SS$ signal.

Complementing the benchmark analysis, a two-dimensional scan over the singlet and heavy Higgs masses reveals the regions of the $m_S$--$m_H$ plane where each search topology is most sensitive. The direct channel $gg\to H\to SS$ dominates across the bulk of the kinematically accessible region, while the channel $gg\to A\to ZH\to ZSS$ and the associated production channel $pp\to tH^\pm\to tW^\pm H\to tW^\pm SS$ are active in more restricted but complementary corners of parameter space, providing $Z$-tagged and $tW$ plus missing transverse energy handles on the same DM final state. The flavor-violating signature $cg \to tH \to tSS$ further enrich the phenomenology and provide additional searches that are 
absent in 2HDM+S models with NFC.

%###################################################################
\section*{Acknowledgments}
We are grateful to George Wei-Shu Hou for many fruitful discussions. 
LMGDLV is supported by the Ministry of Education (Higher Education Sprout Project NTU-113L104022-1), the National Center for Theoretical Sciences~(NCTS) of Taiwan and the Universidad de Sonora FICEN-DADIP project USO315010524.
MK is supported by the National Science and Technology Council of Taiwan under grant No.~114-2639-M-002-006-ASP.
MK would like to thank the NCTS Physics Division for hospitality, where part of this work was carried out.
\appendix 
\counterwithin{figure}{section}
\counterwithin{table}{section}

\section{\texttt{HiggsSignals} Constraints on $c_\gamma$--$\rho_{tt}$ and the $\mathrm{Re}(\rho_{tt})$--$\mathrm{Im}(\rho_{tt})$ Planes} \label{appendix:HS}
\begin{figure}[t]
    \centering
    \includegraphics[width=0.45\linewidth]{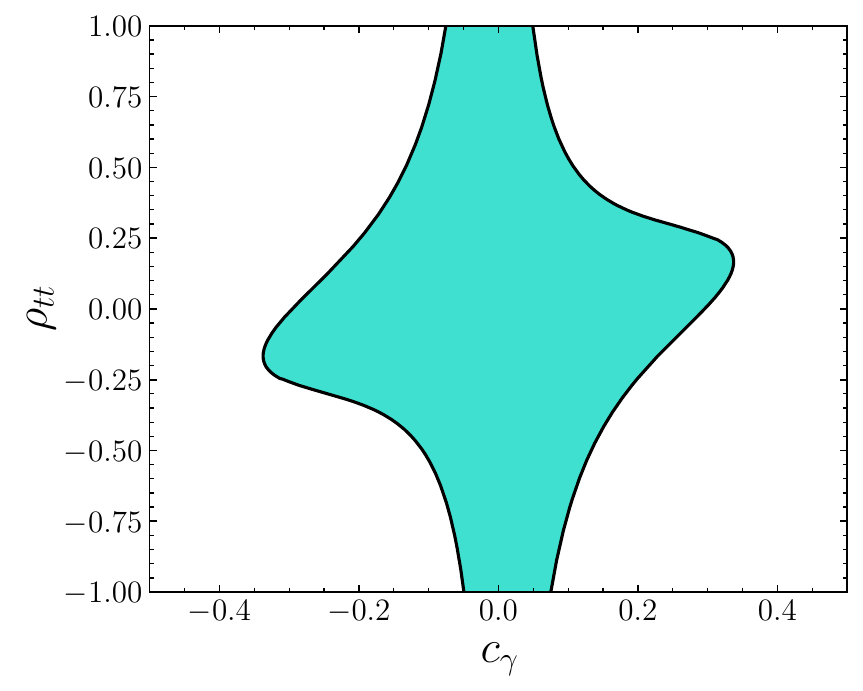}
    \includegraphics[width=0.45\linewidth]{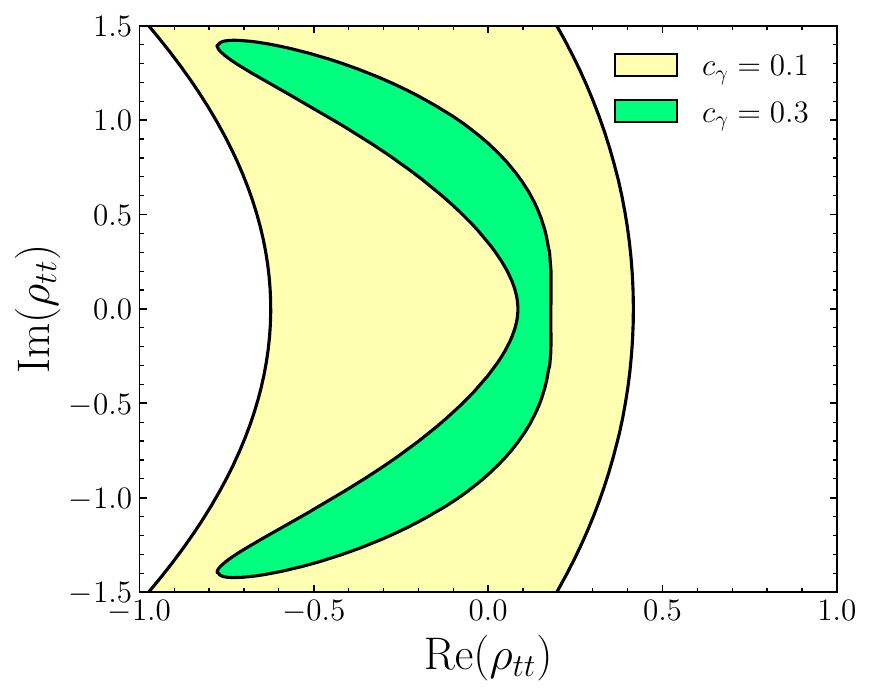}    
    \caption{Allowed regions at $95\%$ CL from \texttt{HiggsSignals}, in the $c_\gamma$--$\rho_{tt}$ plane with $\mathrm{Im}(\rho_{tt}) = 0$~(left) and in the $\mathrm{Re}(\rho_{tt})$--$\mathrm{Im}(\rho_{tt})$ plane~(right).}
    \label{appfig:HS}
\end{figure}

In this appendix, we present constraints on two complementary slices of parameter space: the $c_\gamma$--$\rho_{tt}$ plane with $\mathrm{Im}(\rho_{tt}) = 0$ (left panel of Fig.~\ref{appfig:HS}) and the $\mathrm{Re}(\rho_{tt})$--$\mathrm{Im}(\rho_{tt})$ plane (right panel of Fig.~\ref{appfig:HS}). 
These constraints are derived from the measured signal strengths of the 125 GeV Higgs boson using \texttt{HiggsSignals}, which compares the predicted signal strengths in each channel against the combined LHC Run 1 and Run 2 measurements. 
We refer to Refs.~\cite{Hou:2025fiy,Hou:2018uvr} for further details.

%###################################################################
\section{Effect of Complex Phase of $\rho_{tt}$ on Dark Matter Observables} \label{appendix:phitt}
The presence of a complex phase $\phi$ on the top Yukawa coupling of the second Higgs doublet may have an effect on the dark matter cross sections. 
For the relic density observable, the relevant process is a partial contribution to the dark matter annihilation in the $SS\rightarrow t\bar t$ channel. The norm of the effective coupling of the scalar mediator of this process is 
\begin{equation}
    |\rho_{tt}^{\rm eff}|^2= (|\rho_{tt}|s_\gamma)^2+(\lambda_t c_\gamma)^2+ 2|\rho_{tt}|c_\gamma\lambda_t s_\gamma c_\phi.
    \label{eq:DMannihilationeffectiverho}
\end{equation}
Evidently, when $\phi$ is different from zero this partial contribution to the annihilation cross section is suppressed. However, for $c_\gamma>10^{-2}$, and a value of $|\rho_{tt}|=0.1$ the cross section is dominated in the whole dark matter mass range by bosonic final states. These channels are independent of $\rho_{tt}$, so there is no effect of the phase in this case. For smaller values of $c_\gamma$, the dominant cross section is the two top quark final state, but only for dark matter masses below the $H$ resonance ($2m_S<m_H$). In this case the effect of $\phi$ could be expected to affect the annihilation cross section, but the term that contains it becomes numerically small in eq.~\ref{eq:DMannihilationeffectiverho}. We have checked that varying $\phi$ over its full range induces a variation of less than $5 \%$ on the DM annihilation cross section for the scenarios we study. Because of this, we do not consider further its effect on this observable.

%###################################################################
%\bibliographystyle{unsrt}  
%\bibliographystyle{plain}
\bibliographystyle{apsrev4-1}
\bibliography{references} 

%\begin{thebibliography}{1}

%\end{thebibliography}

\end{document}